\documentclass[journal]{IEEEtran}
\usepackage{amsfonts,epsfig,cite,array,graphicx,amsmath,amssymb,epstopdf}
\usepackage{caption, longtable, makecell}
\usepackage{multirow}
\usepackage{amsmath}
\usepackage{booktabs}
\usepackage{color, colortbl}
\usepackage{adjustbox}
\definecolor{Gray1}{gray}{0.9}
\usepackage{mathtools}
\usepackage{caption}
\usepackage{slashbox}
\usepackage{subcaption}

\usepackage{algorithm}
\usepackage{algpseudocode}
\date{}
\begin{document}

\title{Light-Weight 1-D Convolutional Neural Network Architecture for Mental Task Identification and Classification Based on Single-Channel EEG}
\author{Manali Saini, Udit Satija,~\emph{Member, IEEE},~and~Madhur Deo Upadhayay,~\emph{Member, IEEE}
\thanks{Manali Saini and Madhur Deo Upadhayay are with the Department of Electrical Engineering, Shiv Nadar University, Greater Noida, Udit Satija is with Department of Electrical Engineering, Indian Institute of Technology Patna, Bihar, India. E-mail: ms905@snu.edu.in, udit@iitp.ac.in, madhur\_deo@yahoo.com.}}
\maketitle
\begin{abstract}
Mental task identification and classification using single/limited channel(s) electroencephalogram (EEG) signals in real-time play an important role in the design of portable brain-computer interface (BCI) and neurofeedback (NFB) systems. However, the real-time recorded EEG signals are often contaminated with noises such as ocular artifacts (OAs) and muscle artifacts (MAs), which deteriorate the hand-crafted features extracted from EEG signal, resulting  inadequate identification and classification of mental tasks. Therefore, we investigate the use of recent deep learning techniques which do not require any manual feature extraction or artifact suppression step. In this paper, we propose a light-weight  one-dimensional convolutional neural network (1D-CNN) architecture for mental task identification and classification. The robustness of the proposed architecture is evaluated using artifact-free and artifact-contaminated EEG signals taken from two publicly available databases
(i.e, Keirn and Aunon ($K$) database and EEGMAT ($E$) database) and in-house ($R$) database recorded using single-channel neurosky mindwave mobile 2 (MWM2) EEG headset in performing not only mental/non-mental binary task classification but also different mental/mental multi-tasks classification. Evaluation results demonstrate that the proposed architecture achieves the highest subject-independent classification accuracy of $99.7\%$ and $100\%$ for multi-class classification and pair-wise mental tasks classification respectively in database $K$. Further, the proposed architecture achieves subject-independent classification accuracy of $99\%$ and $98\%$ in database $E$ and the recorded database $R$ respectively. Comparative performance analysis demonstrates that the proposed architecture outperforms existing approaches not only in terms of classification accuracy but also in robustness against artifacts.
\end{abstract}

\vspace{-0.4cm}
\section{Introduction}
Electroencephalogram (EEG) represents the electrical activity of the brain \cite{eeg}. Due to low cost, high temporal resolution and non-invasiveness, EEG is the most commonly used signal in designing neurofeedback (NFB), neural control interface (NCI) and brain computer interface (BCI) systems \cite{pali, spl2020, gupta, wang}. Since portability is one of the critical features for unsupervised mental health monitoring, these systems demand accurate detection of neuronal activities using single/limited channel(s) EEGs \cite{keirn}. It has been shown that EEG signal exhibits different neuronal changes due to various mental activities including, mental tasks and mental stress \cite{tim1, tim2, keirn, btlr}. These changes are induced when the subjects are presented with standardized stress tests, workload tasks, and questionnaires by psychologists \cite{tim1, tim2}. Accurate analysis of these neuronal changes enables identification and classification of different mental tasks which is useful for patients suffering from  motor, cerebral, and behavioral disorders, for example, attention deficit hyperactivity disorder (ADHD) and autism \cite{tim3, tim4}, as well as for healthy persons to improve their concentration and cognitive performance \cite{wang}. Furthermore, the identification and classification of mental tasks from EEG are beneficial for early detection of mental stress and diagnosis of several diseases such as, depression, heart attack, etc \cite{shargie}.

Presently, existing techniques exploit the use of various feature extraction techniques and different machine learning classifiers for mental task identification and classification \cite{zhang, dutta, noshadi, bash, tim1}. Since  single/limited channel(s) EEGs are commonly corrupted with various ocular and muscle artifacts, performance of the hand-crafted features-based mental task identification techniques deteriorates significantly \cite{SatijaSensor2019,SatijaIET2020}. Recently, deep convolutional neural network (CNN) has gained attention due to its ability to extract high level features automatically from the raw data for accurate analysis of different physiological signals \cite{spl2020,confr,ravi}.
Although CNN has been applied on EEG signals for mental workload level classification, there exists no work which utilizes the efficacy of CNN for mental task identification and classification. Furthermore, existing CNN-based mental workload technique \cite{jiao} uses time-frequency representation of EEG in 2D form which demands a complex architecture for learning its 2D features and increases the computational complexity \cite{bash, arxv}. However, real-time NFB system demands low latency in classification process in order to provide timely feedback to the user.

\vspace{-0.4cm}
\subsection{Related work and motivation}
Numerous works have been reported in the literature for the identification and classification of different types of mental tasks from EEG \cite{confr, dutta, zhang}. In  \cite{keirn}, Keirn et al. proposed the use of autoregressive (AR) parameters and band power asymmetry ratios to classify five mental tasks from EEG recordings of five subjects, using Bayes quadratic classifier (BQC). Similar features have been used to train elman neural network (ENN) with resilient backpropagation (RBP) \cite{pali}, and feed forward neural network (NN) \cite{ander}, for classification of different mental tasks. In \cite{dutta}, S. Dutta et al., proposed multivariate AR model based features extracted from intrinsic mode functions in multivariate empirical mode decomposition (MEMD) domain, to classify three cognitive tasks using least squares support vector machine (LS-SVM) classifier. In \cite{noshadi}, modified lempel–Ziv (LZ) complexity has been presented  along with band powers and entropy as features to discriminate five mental tasks using K-nearest neighbour (K-NN) and linear discriminant analysis (LDA) classifiers. Power spectral density based features have been fed to LDA classifier for classification of five mental tasks in six subjects \cite{gupta}. In \cite{lin}, PSD features have also been used along with improved particle swarm optimization (IPSO) based NN classifier to distinguish three mental tasks. Similar features along with statistical features, frequency-bands' power and Higuchi's fractal dimension have been fed to SVM for classification of mental arithmetic tasks in  ten subjects \cite{wang}. In \cite{acmdl}, Alyasseri et al. used subject-independent discrete wavelet transform (DWT) based statistical features along with entropy to classify five mental tasks for seven subjects using artificial neural network (ANN). In \cite{netbio}, EEG signals recorded from 41 subjects during three mental tasks have been classified using subject-independent statistical features and multi-layer perceptron (MLP) kernel based SVM. An immune-feature weighted SVM has been proposed to classify five mental tasks for seven subjects with approximate entropy feature in \cite{guo}.

In \cite{confr}, the EEG waves obtained from discrete wavelet transform of the artifact-removed signal are used as inputs to a 1-D CNN model for discriminating different levels of multimedia learning tasks in $34$ subjects. In \cite{tim1}, Z. Pei et. al. utilized EEG features exhibiting intra-channel and inter-channel information to classify multiple workload tasks with an overall accuracy of $85\%$. In \cite{chen}, P. Zhang, et al., proposed a recurrent three dimensional (3-D) CNN to classify high and low mental workload levels across two tasks, i.e., spatial n-back task and an arithmetic task for twenty subjects. A custom domain adaptation based 3-D CNN with the spatial, spectral and temporal inputs has been used to distinguish four levels of cognitive load from $13$ subjects in \cite{spl2020}. In \cite{jiao}, Jiao et. al., proposed a single-channel spectral- and temporal-spectral-map based CNN model to classify four different levels (low to high) of working memory while recalling some characters from a set shown to $15$ participants.

Most of the existing techniques use subject-dependent hand-crafted features and conventional machine learning approaches for mental task identification and classification \cite{spl2020}. However, these techniques may not generalize across subjects and databases due to high structural and functional changeability between subjects and the non-stationarity  of EEG \cite{spl2020}. This issue can be addressed by the use of deep learning approaches where the features are automatically learnt from the raw data during the training process \cite{craik}. One of the most popular deep learning approaches is CNN, which has been successfully applied for various classification problems related to EEG signals including, seizure detection, emotion recognition, and mental workload level classification \cite{jiao, craik, yuan, spl2, emot, chen}. However, most of these works utilize artifact removal preprocessing step to improve classification accuracy \cite{craik} and/or time-frequency representation of EEG signal as a 2-D or 3-D input to CNN which increases the computational complexity of the complete system \cite{yuan, spl2, emot, chen}. Though, there exists two CNN-based subject-independent mental task classification techniques that classify resting state from multimedia learning task and different mental tasks respectively, they use artifact removal step and DWT representation of the EEG signal \cite{confr} and achieve low accuracy ($70\%$) \cite{arxv1}. Also, the overall CNN architecture becomes computationally complex. Therefore, in this paper, we present a light-weight 1-D CNN architecture which involves few layers and extracts features automatically from the raw EEG signals for accurate identification and classification of mental tasks.

\vspace{-0.35cm}
\subsection{Objective and key contributions}
Literature studies demonstrate that most of the existing hand-crafted feature-based mental task classification techniques have poor subject-independent classification accuracy for both artifact-free and artifact-contaminated EEG signals. Furthermore, use of artifact removal step can alter clinical features of EEG signals even in case of artifact-free EEG signals \cite{SatijaSensor2019, SatijaIET2020}. Existing CNN-based mental task and mental workload classification techniques use complex architecture and input signal in the form of 2D or 3D time-frequency representations of single/multi-channel EEGs \cite{bash}. Therefore, in this paper, we propose a light-weight 1D-CNN architecture for identification and classification of mental tasks from single-channel EEG signal. The significant contributions of this paper are summarized as follows:
\begin{itemize}
\item Investigation of 1D-CNN for automated meaningful feature extraction from the raw single-channel EEG signal.
\item Proposed a low complex subject-independent architecture based on 1D CNN, using only a few layers.
\item Recording of single channel EEG signal using Neurosky mindwave mobile 2 device to examine the performance of the proposed architecture on real-time EEGs.
\item Examination of the classification accuracy of the proposed architecture for not only mental/non-mental task classification but also
several mental/mental multi-tasks classification unlike existing techniques.
\item Examination of the mental task classification accuracy of the proposed architecture using OA- and MA-contaminated  EEG signals.
\item Evaluation of the proposed architecture on three databases, including two publicly available databases and one real-time recorded database.
\end{itemize}
The rest of the paper is organized as follows: Section II describes the databases used in this work. Section III describes the major constituents of the proposed architecture. Section IV discusses the performance results obtained on different EEG signals taken from publicly available databases as well as real-time recorded data. Section V concludes the paper.
\vspace{-0.35cm}
\section{Description of databases}
This section presents a brief description of the three EEG databases used in this work, including two publicly available databases, i.e., Keirn and Aunon database, EEG during mental arithmetic tasks (EEGMAT) database and one recorded database.
\vspace{-0.5cm}
\subsection{Keirn and Aunon database (K)}
This database was recorded by Keirn and Aunon from seven subjects while performing four mental tasks and one baseline/ non-mental task \cite{keirn}.  It consists of both artifact-free and OA-corrupted EEG signals recorded at a sampling rate of $250 Hz$ from six electrode positions, i.e., $C_3,C_4, P_3,P_4,O_1\ and\ O_2$ according to the $10-20$ system, with $A_1$ and $A_2$ as the reference electrodes. Eye blinks (OAs) were also recorded by a separate channel. For each task, recording procedure was performed for $10$ seconds in a sound-controlled booth with dim lighting. Different number of trials were performed by each subject for each task. For example, subjects $1, 3, 4$ and $6$ performed $10$ trials, subjects $2$ and $7$ performed $5$ trials and subject $5$ performed $15$ trials of each task. In this work, we have used all subjects of this database except subject $4$ due to incomplete information available for that subject. The database consists of the signals recorded during the following tasks \cite{keirn}:
\begin{itemize}
\item Baseline task (BT): The subjects were informed to relax and sit idle. No task was performed and the data was recorded in `eyes closed' and `eyes open' position.
\item Multiplication task (MT): The subjects were given a complex multiplication problem to solve without speaking and making any movements.
\item Letter composing task (LT): The subjects were asked to mentally compose a letter to a known person without speaking or making any movement.
\item Geometric figure rotation task (RT): The subjects were presented with a complex $3-D$ figure for $30$ seconds, after which the figure was removed, and they were asked to visualize the same figure being rotated about an axis. The data was recorded during this visualization process.
\item Visual counting task (VT): The subjects were asked to visualize some numbers written in a sequential order on a blackboard, with the previous number being removed before the next number was written.
\end{itemize}
\vspace{-0.35cm}
\subsection{EEG during mental arithmetic tasks (EEGMAT) database (E)}
Database $E$ consists of EEG signals collected from $36$ subjects before and during performance of a mental arithmetic task using Neurocom $23$ channel device \cite{eegmat, physion}. It consists of artifact-free EEG signals of $60s$ duration each, recorded from $F_{P1}$, $F_{P2}$, $F_3$, $F_4$, $F_z$, $F_7$, $F_8$, $C_3$, $C_4$, $C_z$, $P_3$, $P_4$, $P_z$, $O_1$, $O_2$, $T_3$, $T_4$, $T_5$, and $T_6$ electrodes positioned according to the $10-20$ electrode placement system. The sampling frequency was kept at $500 Hz$. Only one trial with $19$ EEG signals was recorded per subject and task. The tasks are as follows:
\begin{itemize}
\item No mental task/ baseline task (BT): The subjects did not perform any mental task and were asked to sit in a relaxed position.
\item Serial subtraction task (ST): Subjects were instructed to perform a serial subtraction task including $4$ digit minuend and $2$ digit subtrahend and communicate the results orally. At the end, the number of subtractions were noted based on the communicated results. A good or a bad quality count was given to each subject based on the results.
\end{itemize}
\vspace{-0.35cm}
\subsection{Recorded database (R)}
\begin{figure}[htbp!]
\centering
\includegraphics[width=5.7cm, height=3.2cm]{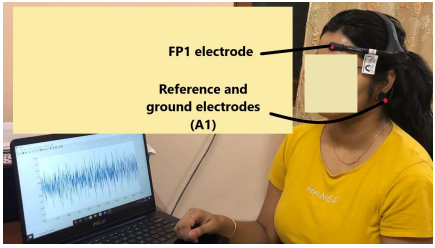}
\caption{\footnotesize{Recording of EEG data of 4$th$ subject from neurosky MWM2 headset while performing the mental task (ST) in `eyes closed' position.}}
\label{fig:fign}
\end{figure}

To evaluate the effectiveness of the feasibility of single-channel EEG data for mental task identification and classification, we recorded in-house EEG signals using twenty subjects during baseline and mental task activity. Details of these subjects have been described in Table \ref{Tab1}. Neurosky mindwave mobile 2 (MWM2) headset was used to record single-channel EEG from $F_{P1}$ position before and during the performance of mental arithmetic task. The headset consists of three electrodes, one for EEG ($F_{P1}$) and other two electrodes  for ground and reference ($A_1$ position) \cite{neuro}, as shown in the recording set up in Fig. \ref{fig:fign}. EEG acquisition has been performed  in a sound controlled laboratory with normal lighting. The inbuilt Thinkgear ASIC (application-specific integrated circuit) module (TGAM) pre-processes the raw signal, i.e., removal of powerline interference (PLI) and MAs. Communication is established between the device and computer using a bluetooth module \cite{neuro}. The data was recorded at a sampling frequency of $512 Hz$ with $12$ bit resolution and analyzed in MATLAB software. Five number of trials of mental and baseline tasks were recorded for each subject and each trial lasted for 10 seconds. Trials are the different sessions of EEG signal recording which were performed during the verbal announcement of another person to `start' and `stop' the session. The following tasks were performed:
\begin{itemize}
\item Baseline task (BT): Subjects were asked  to sit in a relaxed position without making any movement, with ‘eyes open’ and ‘eyes closed’ positions. During this time, the data was labeled as baseline task. This procedure was repeated for all subjects five times, resulting in total five trials/sessions of baseline task per subject.
\item Serial subtraction task (ST): Subjects were instructed  to perform serial subtraction between one 4 digit number (minuend) and other 2 digit number (subtrahend) without speaking and making any movement, in ‘eyes open’ and ‘closed’ position. After the announcement of `start', participants started performing serial subtraction and communicated their subtraction results after the `stop' announcement. Based on their outcome, the number of subtractions performed by each subject was noted by the person. In each trial, different sets of numbers were given for performing ST. To illustrate an example of ST, let the minuend be $4373$ and subtrahend be $59$, then the result after first subtraction: $4373-59=4314$, after second subtraction:  $4373-59-59=4314-59=4254$ and so on. Each participant performed varying number of subtractions depending upon his/her calculation speed.
\end{itemize}
\begin{table}
\centering
\scriptsize
\renewcommand{\tabcolsep}{0.015pc} 
\renewcommand{\arraystretch}{1.0} 
\caption{Database R record details.}
\begin{tabular}{|m{1.25cm}|m{0.33cm}|m{0.33cm}|m{0.33cm}|m{0.33cm}|m{0.33cm}|m{0.33cm}|m{0.33cm}|m{0.33cm}|m{0.33cm}|m{0.33cm}|m{0.33cm}|m{0.33cm}|m{0.33cm}|m{0.33cm}|m{0.33cm}|m{0.33cm}|m{0.33cm}|m{0.33cm}|m{0.33cm}|m{0.33cm}|}

\hline
         Subject number &   1 &2&3&4  & 5 &6&7&8&9&10&11&12&13&14&15&16&17&18&19&20 \\ \hline

        Gender &        M &M&M&F  & F &F&F&M&M&F&F&F&M&M&F&M&M&F&M&M\\ \hline

          Age (Years) &      25 &22&27&28  & 31 &27&30&55&25&28&28&27&25&27&30&32&27&29&26&26 \\ \hline

         Subtractions (number) &    8 &6&7&9  & 6 &5&7&9&8&9&10&9&10&7&8&7&9&8&7&10\\

\hline
\end{tabular}
\label{Tab1}
 \end{table}

\begin{figure}[htbp!]
\centering
\vspace{-0.5cm}
\includegraphics[width=8cm, height=5cm]{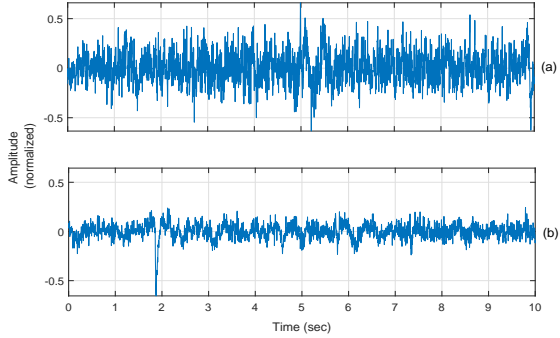}
\vspace{-0.4cm}
\caption{\footnotesize{Recorded EEG signals of 3$rd$ subject (a) and (b) while performing the BT and ST tasks in `eyes closed' position respectively.}}
\label{fig:fign1}
\end{figure}
Fig. \ref{fig:fign1} shows two EEG signals recorded from 3$rd$ subject in `eyes closed' position while performing serial subtraction and baseline tasks. The difference between the amplitudes and frequencies of the two signals is quite observable from the figure. Since the EEG signals have been recorded at different sampling frequencies for all the three databases, all signals have been re-sampled to $500 Hz$ using cubic spline algorithm \cite{SatijaIoT, SatijaJBHI} for adequate classification. 
In this work, we assume that the EEG data is always available and it may be corrupted by in-band ocular artifacts and muscle artifacts, baseline wander and powerline interference. Although use of basic pre-processing
is essential to eliminate flat line, instrumentation noise or raw noise, for which there are well-established techniques, it is out of scope of this work.
\vspace{-0.35cm}
\section{Proposed 1D-CNN architecture}
\begin{figure*}[h]
\centering
\includegraphics[width=18cm, height=8cm]{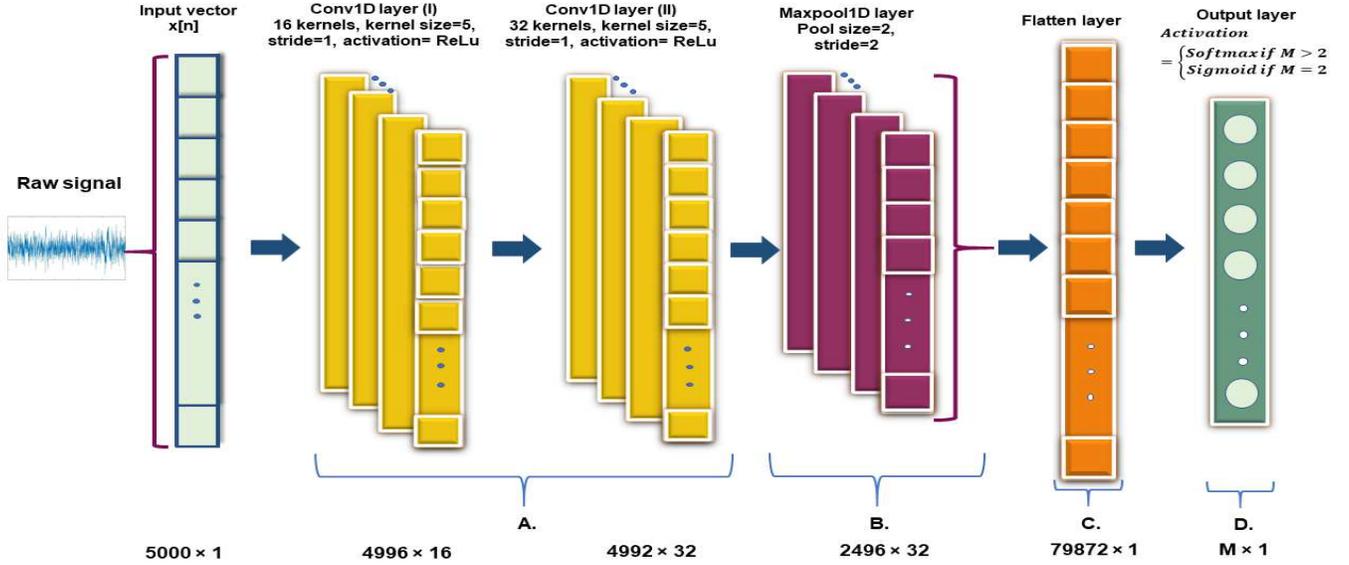}
\vspace{-0.2cm}
\caption{\footnotesize{The proposed CNN architecture for classification of different mental tasks. Note: $M$ denotes the number of classes.}}
\label{fig:fig1}
\end{figure*}

CNN is a popular deep learning approach that has been successfully applied to EEG signal analysis \cite{chen, croce}. It possesses a hierarchical structure in the form of different layers, in which each layer with a specific operation extracts high-level features from the raw input data \cite{access}. In comparison with the conventional fully connected networks, CNN has a tremendous learning capability in extracting more robust features at each layer as well as a good generalization performance \cite{access}. This section presents the proposed 1D-CNN architecture with an input raw single-channel EEG signal denoted as $x[n]$ which is illustrated in Fig. \ref{fig:fig1}. It comprises of two 1D-convolution layers, one 1D-max pooling layer, one flatten layer with dropout and a final dense/ fully connected layer with softmax/ sigmoid activation for classification output.
\vspace{-0.35cm}
\subsection{1D-convolution layer (Conv1D)}
The raw one-dimensional EEG signal (vector) $x[n]$, where, $n=1,2,\ldots N$, is given as an input to the first layer of the CNN architecture, i.e., conv1D, as shown in the Fig. \ref{fig:fig1}. The layer utilizes the following parameters:
\begin{itemize}
\item Filters / Kernels: The filters / kernels produce feature maps by performing convolution with the input signal. The number and size of kernels are crucial for adequately capturing relevant features from the signal.
Let $k[n]$ denote the convolution kernel with size $v$, then the convolution output $c[n]$ can be given as:
\begin{equation}
c[n]=x[n] \ast k[n]=\sum_{m=0}^{v-1} x[m] \cdot k[n-m]
\end{equation}
where, `$\ast$' denotes the convolution operation.
In general, the convolved feature at the output of $l^{th}$ layer can be written as \cite{prez}:
\begin{equation}
c_{i}^{l}=\sigma\left(b_{i}^{l}+\sum_{j} c_{j}^{l-1} \times k_{i j}^{l}\right)
\end{equation}
where, $c_{i}^l$ represents the $i^{th}$ feature in the $l^{th}$ layer; $c_{j}^{l-1}$ denotes the $j^{th}$ feature in the $(l-1)^{th}$ layer;
$k_{i j}^{l}$ represents the kernel linked from $i^{th}$  to $j^{th}$ feature, $b_{i}^{l}$ denotes the bias for this feature and $\sigma$ represents the activation function. 
In the proposed work, two conv1D layers are used. The first convolution layer has $16$ convolution kernels and the second convolution layer has $32$ kernels, each with size $v=5$ and shift / stride = $1$ in both the layers. The output of conv1D layer (I) is given as input to the conv1D layer (II). The length of the output of convolution layer is generally given by $N-v+1$ for stride of $1$, where $N$ is the corresponding input length.
 The convolution operation with aforementioned parameters is illustrated in Fig. \ref{fig:fig2} \cite{prez}, where,\\
\hspace{0.3cm} $c_{1}=k_1 x_1+k_2 x_2+k_3 x_3+k_4 x_4+k_5 x_5$;\\
\hspace{0.3cm} $c_2=k_1 x_2+k_2 x_3+k_3 x_4+k_4 x_5+k_5 x_6$;\\
\vdots\\
Finally,
$c_{N-v+1}=k_1 x_{N-4}+k_2 x_{N-3}+k_3 x_{N-2}+k_4 x_{N-1}+k_5 x_N$.\\
The filter weights, i.e., $k_1 \ldots k_5$, are initialized using the He uniform initializer \cite{hanin} and the bias vector is initialized to all zeros. This operation is performed for each filter in both the layers, hence, there are $16$ outputs of conv1D layer (I) and $32$ outputs of conv1D layer (II). Since $N$ is taken as 5000 in Fig. \ref{fig:fig1}, the output dimensions of conv1D layers (I) and (II) are $4996 \times 16$ and $4992 \times 32$ respectively. Let $lc$ denote the length of final output of the convolution layers, which is $4992$ here.
\begin{figure}[h]
\vspace{-0.3cm}
\hspace{-0.6cm}
\includegraphics[width=9cm, height=3.8cm]{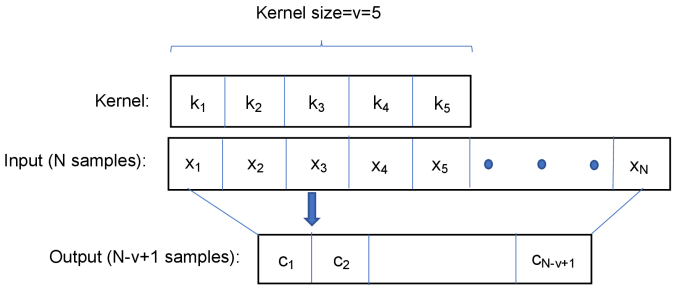}
\caption{\footnotesize{Illustration of convolution operation at conv1D layer.}}
\label{fig:fig2}
\end{figure}\\

\item Activation function ($\sigma$): It plays an important role in capturing the non-linearity of the input signal. Here, rectified linear unit (ReLu) is used as the activation function which is defined as \cite{jiao}:
\begin{equation}
\sigma(c)=\max (0, c)
\end{equation}
\end{itemize}
\vspace{-0.55cm} 
\subsection{1D-max pooling layer (Maxpool1D)}
\begin{figure}[h]
\vspace{-0.3cm}
\includegraphics[width=8cm, height=2.1cm]{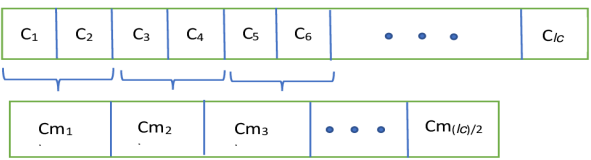}
\caption{\footnotesize{Illustration of max pooling operation with both pool size and stride of two.}}
\label{fig:figmp}
\end{figure}

The output feature maps (convolution outputs, $c$) produced from the conv1D layers are given as an input to the 1D max pooling layer, which reduces the feature map dimension by retaining only the maximum value of feature map in a window/ patch with a specified pool size \cite{jiao}. The window is moved across the feature map with a shift/ stride. The operation of max pooling can be represented as \cite{prez}:
\begin{equation}
c_{h}^{l}=\max _{\forall p \in r_{h}} c_{p}^{l-1}
\end{equation}
where, $r_{h}$ denotes the pooling region with index $h$.\\
In this work, the value of pool size and stride is taken as 2. An illustration of the max pooling operation with these parameters is given in Fig. \ref{fig:figmp}, where,
$c_{m_1}= \max(c_1,c_2)$;\\
$c_{m_2}= \max(c_3,c_4)$;\\
$c_{m_3}= \max(c_5,c_6)$;\\
\vdots\\
$c_{m_{lc/2}}= \max(c_{lc-1},c_{lc})$.\\
Hence, the output of this layer has the dimension of $2496 \times 32$ which can be seen in Fig. \ref{fig:fig1}.
\vspace{-0.4cm} 
\subsection{Flatten layer and dropout}
The flatten layer transforms the input data into a one-dimensional vector, to be fed to the fully connected/ dense layer as shown in the Fig. \ref{fig:fig1}. A dropout parameter is added after the flatten layer, which helps the architecture to generalize well by reducing over fitting during the training process \cite{wu}. This is achieved by randomly setting the activations of some nodes to zero, specified by a dropout rate. In this work, a dropout rate of $0.25$ has been used.
\vspace{-0.45cm}
\subsection{Dense layer for classification}
The flattened output is given as an input to the next layer, i.e., dense/fully connected layer which produces the classification output with dimension $M \times 1$, where $M$ denotes the number of classes. In general, the layer operation can be represented as:
\begin{equation}
output= \sigma(<input, w_d> + ~b_d)
\end{equation}
where, $<input, w_d>$ denotes the dot product between weight vector $w_d$ used in this layer and the input, $b_d$ represents the bias vector for this layer and $\sigma$ is the activation function.
In this work, we use both sigmoid and softmax activation for  binary and multi-class classification respectively. The sigmoid activation function is given by \cite{elfw}:
\begin{equation}
\sigma(z)=\frac{1}{1+e^{-z}}
\end{equation}
This function produces the binary output as the probability value for binary classification, based on which the class label is either `0' or `1'. Also, the softmax activation function can be given as \cite{prez}:
\begin{equation}
\operatorname{softmax}(\textbf{z})_{i}=p_{i}=\frac{\exp \left(z_{i}\right)}{\sum_{j=1}^{M} \exp \left(z_{j}\right)}
\end{equation}
where, $z_i$ represents the $i^{th}$ element of the output vector of previous layer $\textbf{z}$. The numerator is normalized by the sum of all exponential terms from 1 to $M$ to bring the value of $p_i$ between 0 and 1. This layer produces the categorical class labels for multi-class classification. In this work, no bias vector has been used for this layer and the weights are initialized using the glorot uniform initializer \cite{hanin}.
\vspace{-0.45cm}
\section{Results and Discussions}
In this section, the performance of the proposed architecture is evaluated using different artifact-free and artifactual EEG signals taken from publicly available databases and our recorded database.
\vspace{-0.45cm}

\subsection{Performance metrics and training parameters}
The performance of the proposed architecture is assessed in terms of following performance metrics \cite{guo}:
\begin{equation}
\text{Classification accuracy (Accuracy)}= \frac{TP+TN}{TP+TN+FP+FN}
\end{equation}
Here, TP denotes true positives which is the number of cases where the actual positive class is correctly predicted to be positive by the model. TN denotes true negatives which is the number of cases where the actual negative class is correctly predicted to be negative by the model. FP denotes false positives which is the number of cases where the actual negative class is incorrectly predicted to be positive by the model. FN denotes false negatives which is the number of cases where the actual positive class is incorrectly predicted to be negative by the model. 
For example, in binary classification problem of datasets E and R, we have labeled the non-mental (BT) task as negative and mental (ST) task as positive. If the model predicts a BT task correctly, then it is a true negative. Otherwise, if the model predicts it as an ST task, then it is a false positive. Similarly, if the model predicts an ST task correctly, then it is a true positive, and if it predicts it as BT task, then it is a false negative. Similar interpretation can be drawn for the case of multi-class classification.
\begin{equation}
\text{Precision (PRC)} =T P /(T P+F P)
\end{equation}
\begin{equation}
\text{Recall (RCL)} =T P /(T P+F N)
\end{equation}
\begin{equation}
\begin{aligned} F\text {1 score}=& 2 \times \text { PRC } \times \text { RCL } \\ & /(\text { PRC }+\text { RCL }) \end{aligned}
\end{equation}
\vspace{-0cm}

The performance is evaluated through training and testing of the proposed architecture for the identification and classification of mental tasks. Similar to existing works \cite{gupta, lin}, EEG signal of $10sec$ duration has been used as an input to the first layer of the model. We perform both binary classification and multi-class classification using the same architecture. In this work,  following tasks have been classified: BT-MT, BT-LT, BT-RT, BT-VT, MT-LT, MT-RT, MT-VT, LT-RT, LT-VT, RT-VT, BT-ST and BT-MT-LT-RT-VT from all the three databases. To evaluate the performance of the proposed architecture, 80\% of the data is chosen for training and 20\% for testing. $20\%$ data for testing is further split into $10\%$ each for testing and validation. Since different subjects have varying number of trials recorded on separate timings/days, $80\%$ of the trials have been selected randomly for training and rest $20\%$ for testing.
For training, a batch size of $50$, and $20$ number of epochs have been used along with the Adam learning algorithm with a learning rate of $0.001$. Ten fold cross-validation has been performed for all the three databases. Binary cross entropy and categorical cross entropy are used as loss functions for binary and multi-class classification respectively. These functions are defined as \cite{ig}:
\begin{equation}
\text{Binary cross entropy}=-(y \log (p)+(1-y) \log (1-p))
\end{equation}
\begin{equation}
\text{Categorical cross entropy}=-\sum_{c=1}^{M} y_{o, c} \log \left(p_{o, c}\right)
\end{equation}
where, $log$ represents natural log, $y$ represents binary indicator (0 or 1) if class label $c$ is the correct classification for the observation $o$, $p$ represents the predicted probability that the observation $o$ is of class label $c$, $M$ represents the number of classes.
\vspace{-0.45cm}

\subsection{Performance analysis}
\begin{figure}
\includegraphics[width=9cm, height=3.2cm]{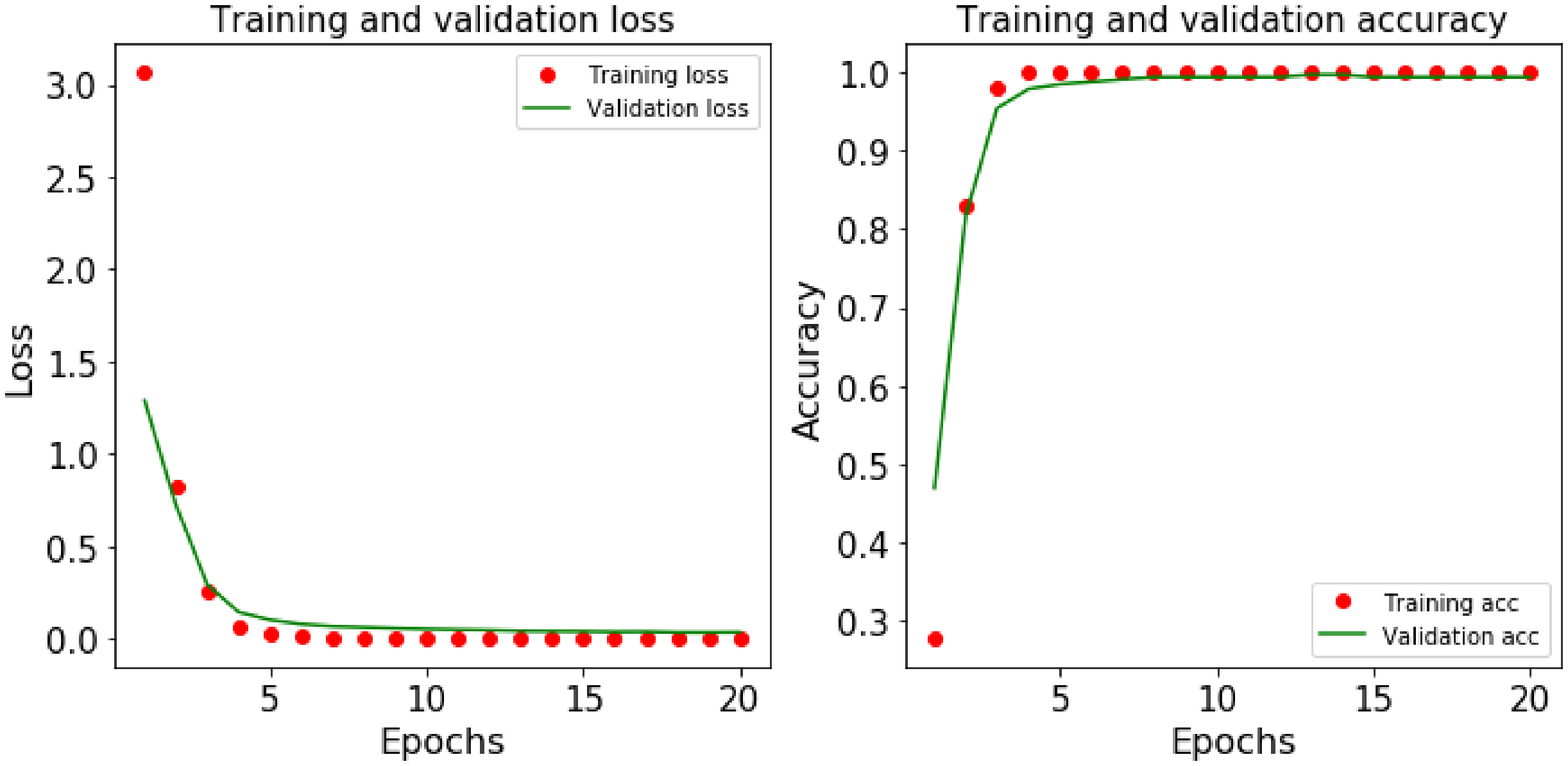}
\vspace{-0.5cm}
\caption{\footnotesize{Training curves for multi-task classification (BT-MT-LT-RT-VT) in database $K$.}}
\label{fig:figt1}
\end{figure}

\begin{figure*}
\includegraphics[width=17.5cm, height=13.5cm]{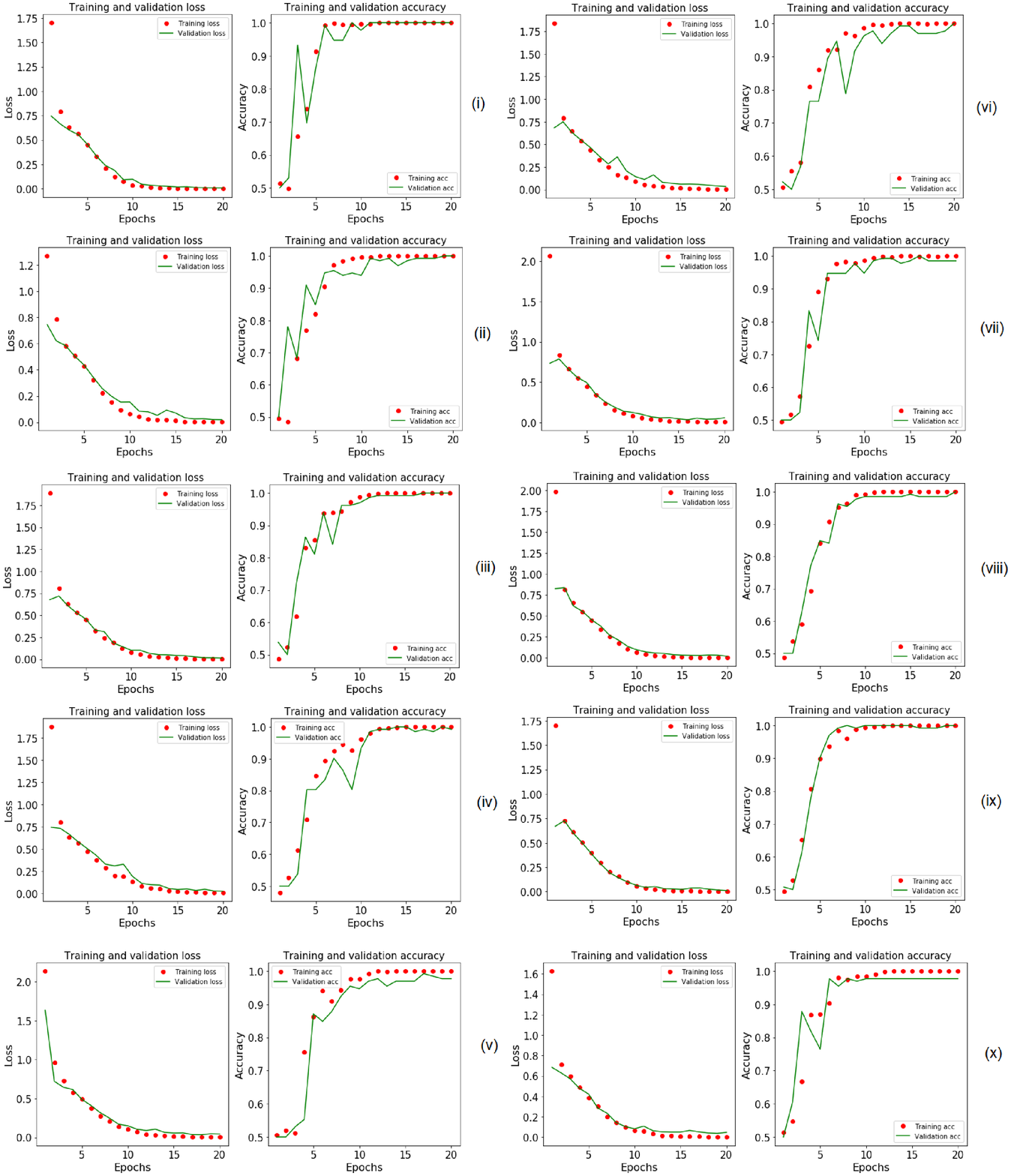}
\caption{\footnotesize{Training curves for pair-wise mental task classification in database $K$, (i) BT-MT, (ii) BT-RT, (iii) MT-LT, (iv) MT-VT, (v) LT-VT, (vi) BT-LT, (vii) BT-VT, (viii) MT-RT, (ix) LT-RT and (x) RT-VT.}}
\label{fig:figt2}
\end{figure*}

In this section, we demonstrate the classification performance results of the proposed architecture. Figs. \ref{fig:figt1}, \ref{fig:figt2} depict the training curves with respect to validation and training loss, and validation and training accuracy for the proposed architecture which demonstrate the learning process for the multi-task classification and pair-wise mental task classification in the database $K$ respectively. It can be observed from the curves that the proposed architecture has learnt from the given data in few epochs and does not over-fit. Table \ref{tab:tab1} depicts the performance of the proposed architecture in terms of aforementioned performance metrics for all classification tasks and databases. It can be observed from the table that a subject-independent accuracy of $100 \%$ has been achieved for the following task pairs: BT-MT, BT-LT, BT-RT, MT-LT, MT-RT, and LT-RT, of the database $K$. It means that the these tasks are accurately classified. It can be observed from the table that the other metrics PRC, RCL are also equal to $1$ for these task pairs, which implies that there are no false positives and false negatives respectively. This results in a perfect balance between PRC and RCL, as observed from the F1 score values (which are also $1$) for these tasks. Further, an overall average subject-independent accuracy of $99.7\%$ has been achieved for the case of multi-class classification for this database. For databases $E$ and $R$, proposed architecture achieves an average subject-independent accuracy of $99\%$ and $98\%$ for the classification of BT-ST task pair. 

\begin{table*}[!htbp]
  \centering
\caption{\small{Mental task classification performance of the proposed architecture for all databases (Mean(std) over subjects ($10 s$ data))}.}
  \newcommand{\cc}[1]{\multicolumn{1}{c}{#1}}
\renewcommand{\tabcolsep}{0.18pc} 
\renewcommand{\arraystretch}{1.2} 
\scalebox{0.9}{
    \begin{tabular}{|c|l|l|l|l|l|l|l|l|l|l|l|l|l|l|}
    \hline
  \multicolumn{1}{|c|}{$\textbf{Database}$}   & \multicolumn{11}{c|}{$\textbf{K}$}                                                               & \multicolumn{1}{c|}{$\textbf{E}$} & \multicolumn{1}{c|}{$\textbf{R}$}  & \multicolumn{1}{c|}{$\textbf{N}$}\\
    \hline
      \backslashbox{\textbf{Metrics}}{\textbf{Task}}  & \multicolumn{1}{c|}{\textbf{BT-MT}} & \multicolumn{1}{c|}{\textbf{BT-LT}} & \multicolumn{1}{c|}{\textbf{BT-RT}} & \multicolumn{1}{c|}{\textbf{BT-VT}} & \multicolumn{1}{c|}{\textbf{MT-LT}} & \multicolumn{1}{c|}{\textbf{MT-RT}} & \multicolumn{1}{c|}{\textbf{MT-VT}} & \multicolumn{1}{c|}{\textbf{LT-RT}} & \multicolumn{1}{c|}{\textbf{LT-VT}} & \multicolumn{1}{c|}{\textbf{RT-VT}} & \multicolumn{1}{c|}{\textbf{BT-MT-RT-LT-VT}} & \multicolumn{1}{c|}{\textbf{BT-ST}} & \multicolumn{1}{c|}{\textbf{BT-ST}}& \multicolumn{1}{c|}{\textbf{BT-ST}} \\
    \hline
     \textbf{Accuracy} & 1 (0) & 1 (0) & 1 (0) & 0.99 (0.12) & 1 (0) & 1 (0) & 0.99 (0.12) & 1 (0) & 0.98 (0.40) & 0.98 (0.21) & 0.997 (0.11) & 0.99 (0.11)& 0.98 (0.22) &  0.99 (0.16) \\
    \textbf{PRC} & 1 (0) & 1 (0) & 1 (0) & 0.99 (0.20) & 1 (0) & 1 (0) & 0.99 (0.11) & 1 (0) & 0.98 (0.25) & 0.98 (0.20) & 0.99 (0.18) & 0.99 (0.15)& 0.98 (0.27) & 0.99 (0.22) \\
    \textbf{RCL} & 1 (0) & 1 (0) & 1 (0) & 0.98 (0.11) & 1 (0) & 1 (0) & 0.99 (0.11) & 1 (0) & 0.98 (0.22) & 0.98 (0.11) & 0.99 (0.10) & 0.99 (0.17) & 0.98 (0.21) &0.99 (0.21) \\
    \textbf{F1}  & 1 (0) & 1 (0) & 1 (0) & 0.98 (0.10) & 1 (0) & 1 (0) & 0.99 (0.11) & 1 (0) & 0.98 (0.32) & 0.98 (0.12) & 0.99 (0.16) & 0.99 (0.10) &0.98 (0.20) & 0.99 (0.17) \\
    \hline
    \end{tabular}}
  \label{tab:tab1}%
\end{table*}%
\vspace{-0.45cm}
\subsection{Impact of EEG processing length and number of conv1D layers}
\begin{figure}[ht!]
\hspace{-0.75cm}
\begin{subfigure}{.33\textwidth}
  \raggedright
  \includegraphics[width=5cm,height=5cm]{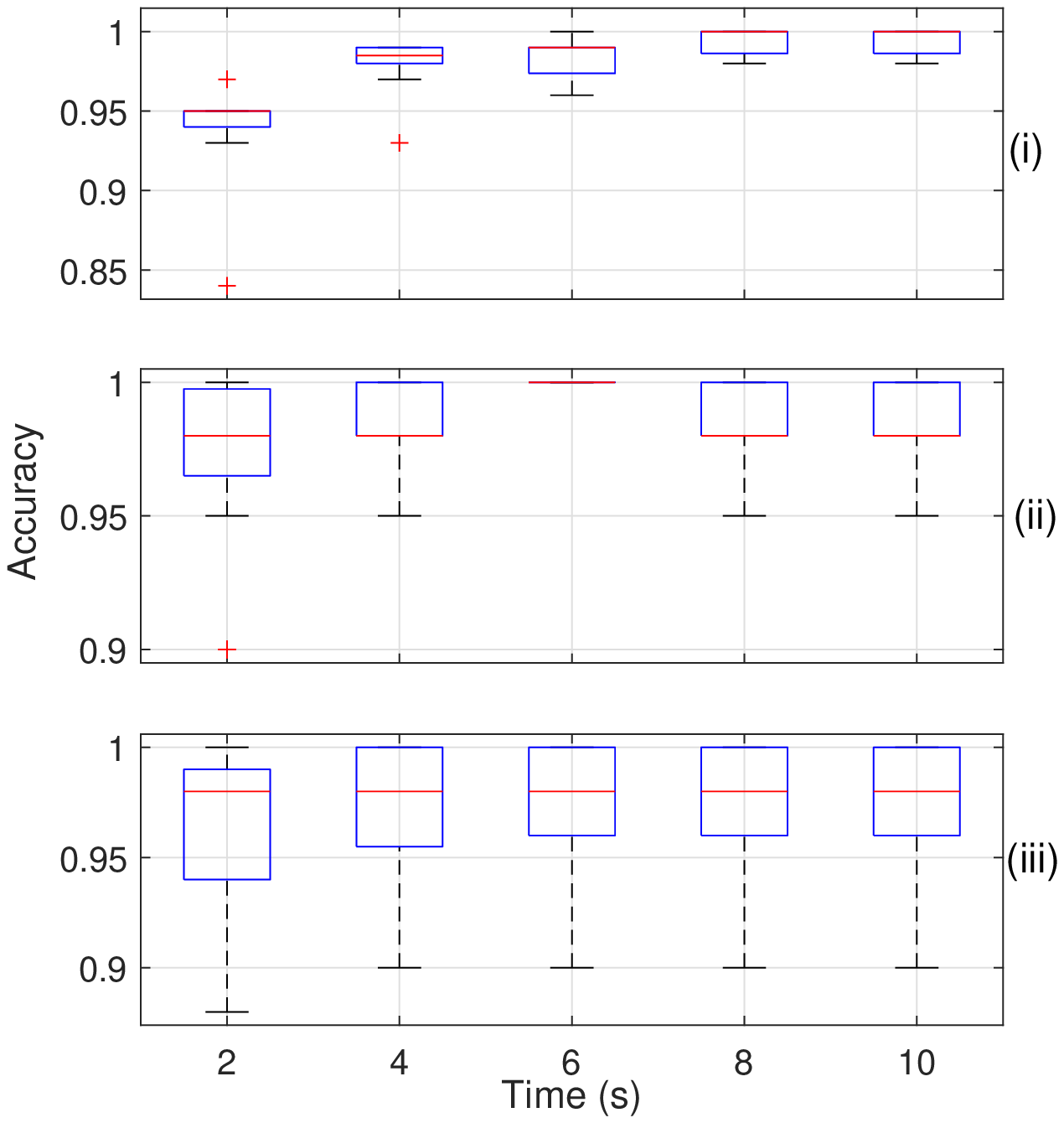}
  \hspace{-1cm}
  \vspace{-0.2cm}
\caption{\footnotesize{}}
  \label{fig:sfig1}
\end{subfigure}%
\begin{subfigure}{.32\textwidth}
  \raggedright
  \hspace{-1.4cm}
  \includegraphics[width=5cm,height=5cm]{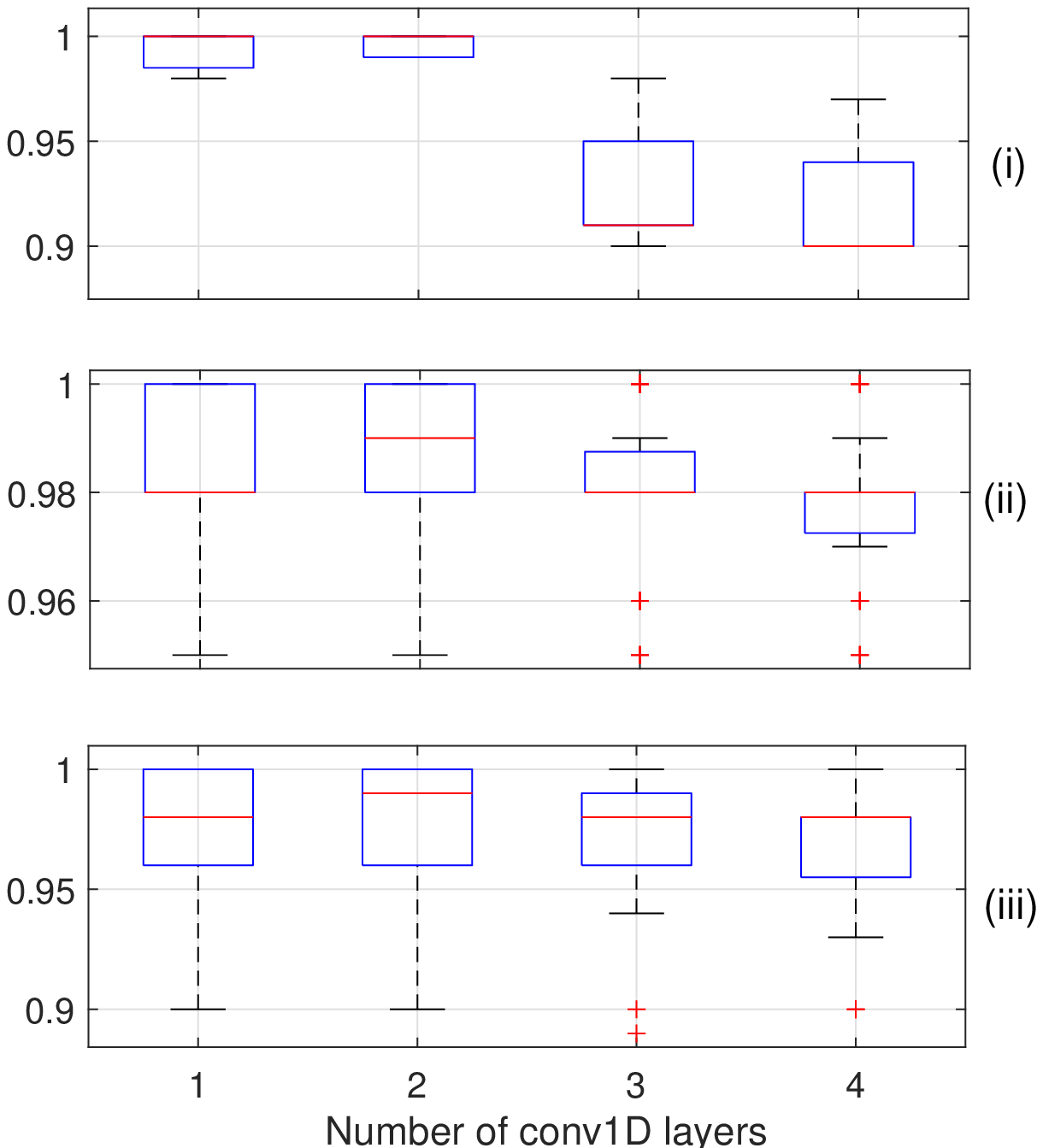}
  \vspace{-0.2cm}
  \caption{\footnotesize{}}
  \label{fig:sfig2}
\end{subfigure}
\vspace{-0.25cm}
\caption{\footnotesize{Illustrates average classification accuracies for (a) varying lengths of input signal and (b) varying number of conv1D layers in: (i) database $K$, where the accuracies are shown with respect to pair-wise mental task classification, (ii) and (iii) databases $E$ and $R$ with respect to all the subjects respectively.} }
\label{fig:figvariation}
\end{figure}

For assessing the sensitivity performance of the proposed architecture with respect to the processing length of the input signal and the number of conv1D layers, different input lengths ranging from $2s$-$10s$ with a step of $2s$ and varying number of conv1D layers have been used. It is quite observable from the Fig. \ref{fig:sfig1} that the proposed architecture is not significantly variant to the input signal length. Even for shorter durations of processing length, the proposed architecture achieves similar accuracies for different mental tasks classifications. This is useful for the systems where quick response is needed, for example, BCI and neurofeedback systems. Further, it can be observed from the Fig. \ref{fig:sfig2} that the mental tasks classification accuracy is higher for two conv1D layers as compared to one or more than two layers in the proposed architecture. Therefore, two conv1D layers are optimal in the proposed architecture for the classification of mental tasks.
\vspace{-0.35cm}
\subsection{Robustness evaluation under different artifacts}
\begin{figure}[h]
\includegraphics[width=9cm, height=3.1cm]{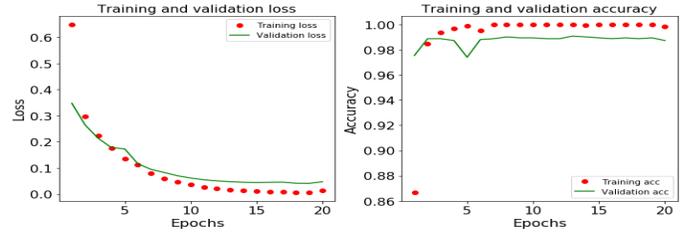}
\caption{\footnotesize{Training curves for BT-ST task classification in created artifact-contaminated database $N$.}}
\label{fig:figt3}
\end{figure}

\begin{figure}[h]
\vspace{-0.65cm}
\includegraphics[width=8cm, height=4.2cm]{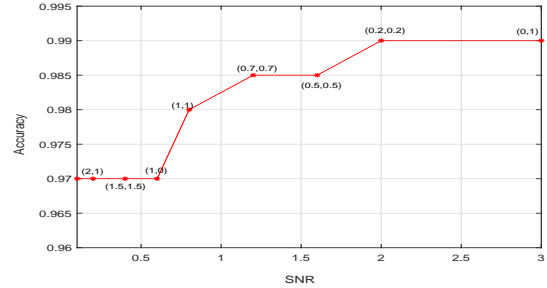}
\vspace{-0.35cm}
\caption{\footnotesize{Illustrates the classification accuracies at different values of ($\lambda$,$\beta$) and corresponding SNRs for the realized contaminated signals in database $N$.}}
\label{fig:figsnr3}
\end{figure}

\vspace{-0.35cm}
In this subsection, we evaluate the robustness of the proposed architecture under different artifacts in the EEG signal. As mentioned earlier, database $E$ contains artifact-free EEG signals and databases $K$ and $R$ consist of EEG signals contaminated with ocular artifacts and muscle artifacts. Hence, in order to evaluate the robustness of the proposed architecture on a large pool of contaminated EEG data, we create a new database namely '$N$' by including only artifact-contaminated EEG signals from database $R$ and all artificially contaminated EEG signals of database $E$ after adding OAs and MAs taken from existing publicly available databases. These OAs and MAs are taken from Mendeley database \cite{mendeley} and MIT BIH polysomnographic database \cite{physion}, and examples of electromyogram database \cite{physion} and cerebral vasoregulation in elderly with stroke (CVES) database \cite{physion} respectively. Various realizations of the contaminated data are generated by randomly adding OAs and MAs separately, as well as both together, to each signal of the database $E$, after re-sampling all OAs and MAs signals to the rate of $500 Hz$. Let $x_{cl}[n]$ denote an EEG signal from the database $E$, $y_o[n]$ denote an OA signal, and $y_m[n]$ denote a MA signal. Then, this process can be summarized as follows:
$x_o[n]= x_{cl}[n]+\lambda \cdot (y_o[n])$, 
$x_m[n]= x_{cl}[n]+\beta \cdot (y_m[n])$, and
$x_{om}[n]= x_{cl}[n]+\lambda \cdot (y_o[n])+\beta \cdot (y_m[n])$.
Here, $x_o[n]$ represents OA-contaminated signal, $x_m[n]$ denotes MA-contaminated signal, and $x_{om}[n]$ represents combined OA-MA-contaminated signal. $\lambda$ is a parameter which denotes the contribution of OAs and $\beta$ denotes the contribution of MAs. These parameters alter the signal to noise ratio (SNR) of the realized signal \cite{ccam}. For example, the SNR for OA contaminated signal can be obtained as \cite{ccam}:
$\mathrm{SNR_{(x_o[n])}}=\frac{\sqrt{\frac{1}{n}\sum_{n} x^2_{cl}[n]}}{\sqrt{\frac{1}{n}\sum_n(\lambda \cdot y_o[n])^2}}$.
Similarly, the SNR values for other realizations of contaminated data can be obtained. In this work, the value of $\lambda$ and $\beta$ is chosen as 1 such that the EEG signal is fairly contaminated by artifacts. Fig. \ref{fig:figt3} depicts the training curve with respect to validation and training loss, and validation and training accuracy for the proposed architecture which demonstrate the learning process for classifying BT-ST task in the created database $N$. In order to demonstrate the robustness of the proposed architecture, mental classification accuracy is computed for contaminated EEG signals with different SNR values ranging $0.4$ - $3$ which are calculated based on different values of $\lambda$ and $\beta$. Fig. \ref{fig:figsnr3} depicts the average classification accuracy at different SNRs which demonstrates the robustness of the proposed architecture in classifying mental and baseline tasks for both artifact-free and artifact-contaminated EEG signals. Also, the average performance of the proposed architecture in terms of performance metrics is demonstrated in Table \ref{tab:tab1} for the created database $N$. As mentioned earlier, the average accuracy is computed for the value of $\lambda$ = $\beta$ = 1 and corresponding SNR value of 0.8, as shown in Fig. \ref{fig:figsnr3}. It can be observed from the table that the proposed architecture achieves an average accuracy of $99\%$ for the artifact-contaminated database $N$.
\vspace{-0.3cm}
\subsection{Performance comparison}
In this subsection, the supremacy of the proposed architecture for mental task classification is analyzed with respect to the other existing mental task classification techniques. Table \ref{tab:tabc} demonstrates the overall comparison of the proposed architecture with the existing mental task classification techniques. It can be observed from the table that the proposed architecture not only outperforms existing approaches in terms of accuracy for mental tasks classification but also does not use an artifact suppression/ removal step before training unlike existing approaches. Furthermore, it can be seen from the table that the proposed architecture can accurately classify not only pair-wise mental/baseline task but also five multi-tasks simultaneously unlike existing approaches.
\begin{figure*}[h]
\includegraphics[width=18.2cm,height=7cm]{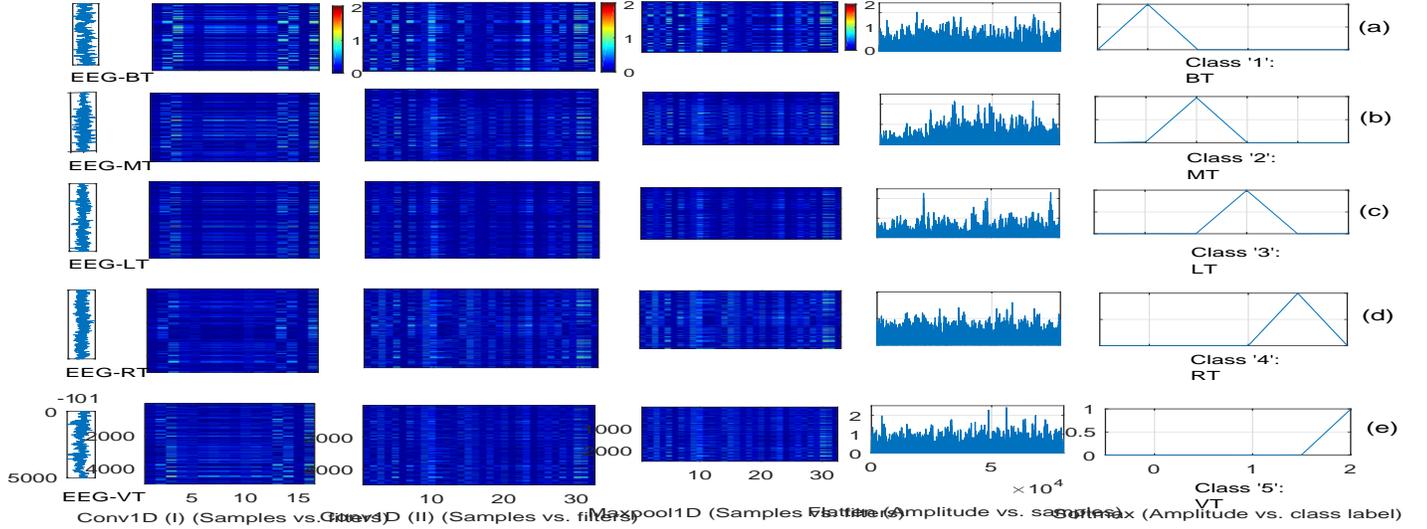}
\caption{Feature/activation maps of 1D-CNN layers for signals taken from database $K$, acquired during different mental tasks.}
\label{fig:figf1}
\end{figure*}
\begin{figure*}
\vspace{-0.25cm}
\includegraphics[width=18.2cm,height=4cm]{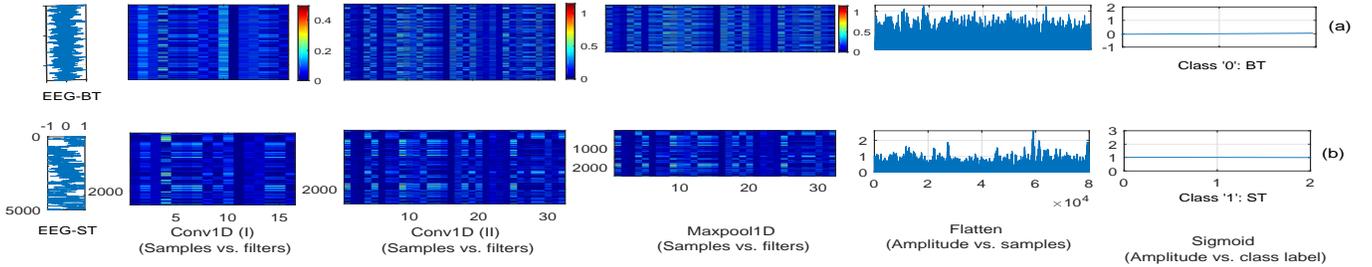}
\vspace{-0.8cm}
\caption{Feature/activation maps of 1D-CNN layers for signals taken from database $N$ acquired during BT and ST.}
\label{fig:figf2}
\end{figure*}
\begin{table}[!htbp]
\centering
\vspace{-0.35cm}
\caption{\small{Comparison of the proposed architecture with the existing approaches.}}
\newcommand{\cc}[1]{\multicolumn{1}{c}{#1}}
\renewcommand{\tabcolsep}{0.2pc} 
\renewcommand{\arraystretch}{1.2} 
\scalebox{0.5}{
\begin{tabular}{|m{2.2cm}|m{2.2cm}|m{2.4cm}|m{1.5cm}|m{1.5cm}|m{1.2cm}|m{1.6cm}|}
\hline

{\bf Method (Author)} & {\bf Database} & {\bf Channel locations (Number)} & {\bf Subjects (Number)} & {\bf Artifact removal/ rejection} & {\bf Length used} & {\bf Accuracy (\%)} \\
\hline
BQC (Keirn and Aunon) \cite{keirn} & K &  $C_3$, $C_4$, $P_3$, $P_4$, $O_1$, $O_2$ (6) &          5 &        Y &         2s & 85-95 \\
\hline
FF-NN (Anderson et. al.) \cite{ander} & K & $C_3$, $C_4$, $P_3$, $P_4$, $O_1$, $O_2$ (6) &          4 &        Y &        10s &     38-71   \\
\hline
IF-SVM (Guo et. al.) \cite{guo} & K &         $C_3$, $C_4$, $P_3$, $P_4$, $O_1$, $O_2$ (6) &          7 &        Y &        10s &    88-98  \\
\hline
LDA, QDA, SVM (Gupta et. al.) \cite{gupta} & K &          $C_3$, $C_4$, $P_3$, $P_4$, $O_1$, $O_2$ (6)&          6 &         N &        10s & 85  \\
\hline
KNN, LDA (Noshadi et. al.) \cite{noshadi} & K &        $C_3$, $C_4$, $P_3$, $P_4$, $O_1$, $O_2$ (6) &          7 &        Y &        10s  & 97 \\
\hline
ENN-RBP (Palaniappan) \cite{pali} & K &         $C_3$, $C_4$, $P_3$, $P_4$, $O_1$, $O_2$ (6) 6 &          4 &        Y &        10s & 80-94  \\
\hline
LS-SVM (Dutta et. al.) \cite{dutta} & K &         $C_3$, $C_4$, $P_3$, $P_4$, $O_1$, $O_2$ (6) 6 &          3 &         N &        10s  & 94          \\
\hline
BP-MLANN (Alyasseri et. al.) \cite{acmdl} & K &     $C_3$, $C_4$, $P_3$, $P_4$, $O_1$, $O_2$ (6) &          7 &        Y &        10s & 78-87    \\
\hline
SVM-MLP (Anand et. al.) \cite{netbio} & Self recorded &  $F_3$, $F_Z$, $F_4$, $C_3$, $C_Z$, $C_4$, $P_3$, $P_{Oz}$, $P_4$      (9) &         41 &         N &      5 min &  73   \\
\hline
IPSO-NN (Lin et. al.) \cite{lin} & IDIAP BCI-III &    $C_3$, $C_z$, $C_4$, $C_{P1}$, $C_{P2},$ $P_3$, $P_z$, $P_4$ (8)        &          3 &         N &      4 min &  69       \\
\hline
SVM-RBF (Wang et. al.) \cite{wang} & Self recorded &   $A_{F3}$, $F_7$, $F_3$, $F_{C5}$, $T_7$, $P_7$, $O_1$, $O_2$, $P_8$, $T_8$, $F_{C6}$, $F_4$, $F_8$, $A_{F4}$      (14) &         10 &        Y &      2 min  & 97   \\
\hline
Random forest (Z. Pei et. al.) \cite{tim1} & Self recorded &  62 EEG channels &  7 &  Y & 2s  & 85   \\
\hline

\multicolumn{ 1}{|l|}{Proposed} & K &        $C_3$, $C_4$, $P_3$, $P_4$, $O_1$, $O_2$ (6)&          6 &         N &        10s & 99.7 (MT), 99 (TP) \\ \cline{2-7}

\multicolumn{ 1}{|l|}{} &     E &    $F_{P1}$, $F_{P2}$, $F_3$, $F_4$, $F_z$, $F_7$, $F_8$, $C_3$, $C_4$, $C_z$, $P_3$, $P_4$, $P_z$, $O_1$, $O_2$, $T_3$, $T_4$, $T_5$, $T_6$   (19) &         36 &         N &        10s  & 99   \\ \cline{2-7}

\multicolumn{ 1}{|l|}{} & R &         $F_{P1}$ (1) &          20 &         N &        10s &  98  \\

\hline
\end{tabular}}
\label{tab:tabc}
\vspace{0.1cm}\\
\tiny{BQC: Bayesian quadratic classifier, FF-NN: Feed forward neural network, IF-SVM: Immune feature weighted SVM, QDA: Quadratic discriminant analysis, KNN: K nearest neighbor, BP-MLANN: Backpropagation-multilayer artificial neural network, SVM-RBF: SVM-radial basis function, Y: Yes, N: No.}
\end{table}
To demonstrate the learning process through both the layers, the feature/activation maps are produced in the proposed 1D-CNN architecture. The feature maps are extracted as filter weights for conv1D and maxpool1D layers, based on which the outputs of flatten layer and softmax layer correspond to the particular class (mental task). Figs. \ref{fig:figf1}-\ref{fig:figf2} depict the features extracted by the proposed 1D-CNN from EEG signals with baseline and different mental tasks taken from database $K$ and $N$ respectively.
From the figures, it can be observed that for each class, the feature maps have inhibitory (small) or excitatory (large) weights for different EEG signals which is illustrated by lighter squares in the Conv1D and Maxpool1D  filters. From \ref{fig:figf1} (a), (b) it can be seen that more filter weights are excitatory in the case of BT, as compared to MT. Also, for the case of RT as shown in Fig.\ref{fig:figf1}(d), the activations in maxpool1D are higher for all filters, indicating the brain is in attention state throughout the duration, which is true since the task involves memorizing the figure and visualization of the figure rotation. Similar interpretation can be drawn from the feature maps of other mental tasks. Further, it can be observed from the Fig.\ref{fig:figf2} (a), (b) that the activations for BT are higher and uniform for all layers, while for ST, activations are higher in only certain locations. These uniform activations in BT indicate the subject's relaxed mental state as against random patterns of activations in ST which indicate that the subject has concentrated on some mental activity. Therefore, these features can be mapped to the neurophysiology of the brain. Our proposed CNN architecture has adequately learned discriminatory feature maps for baseline and different mental task classification as depicted by different activation weights which yield high accuracy.
To demonstrate the subject adaption of the proposed architecture, we train the proposed model on all signals from EEGMAT and test on the subjects from recorded data. A classification accuracy of $97\%$ has been achieved in this case, which is quite high and indicates the subject-adaptability of proposed architecture. While, the existing features fail in capturing the subject-adaptability, as seen from the performance comparison results. In the future direction, we will analyze the performance of the proposed architecture in case of missing EEG samples and implement the proposed architecture on a real-time embedded processor to determine the real-time latency and power consumption in mental task classification.
\vspace{-0.45cm}
\section{Conclusion}
\vspace{-0.15cm}
In the proposed work, a light-weight one-dimensional convolutional neural network (1D-CNN) architecture is proposed for mental task identification and classification.
The proposed architecture consists of a few layer network which does not require any manual feature extraction or artifact suppression step. The proposed
architecture is evaluated using two publicly available databases (i.e, Keirn and Aunon ($K$) database and EEGMAT ($E$) database) and in-house
database ($R$) recorded using single-channel neurosky mindwave mobile 2 (MWM2) EEG headset for performing mental/baseline binary classification
and mental/mental multi-tasks classification.  The proposed architecture achieves classification accuracies of $100 \%$ for the following binary task pairs: BT-MT, BT-LT, BT-RT, MT-LT, MT-RT, and LT-RT, in the database $K$. Further, the proposed architecture achieves an overall average accuracy of $99.7\%$ for multi-class classification in database $K$, and $99\%$ and $98\%$ for the classification of BT-ST task pair in the databases $E$ and $R$ respectively. Comparative performance results show that the proposed architecture outperforms existing approaches not only in terms of classification accuracy but also in robustness against artifacts. Further, the proposed architecture provides good classification accuracy for shorter processing length of EEG signals which makes it suitable for BCI systems with neurofeedback.


\vspace{-0.35cm}
\begin{thebibliography}{1}
\vspace{-0.25cm}
\bibitem{eeg}
D.P. Subha et. al., ``EEG signal analysis: a survey," \textit{J Med Syst}, vol. 34, pp. 195–212, 2010.

\bibitem{pali}
R. Palaniappan, ``Utilizing gamma band to improve mental task based brain-computer interface design," \textit{IEEE Transactions on Neural Systems and Rehabilitation Engineering}, vol. 14, no.3, pp. 299-303, 2006.

\bibitem{spl2020}
M. Jimnez-Guarneros and P. Gomez-Gil, ``Custom domain adaptation: a new method for cross-subject, EEG-based cognitive load recognition," \textit{in IEEE Signal Processing Letters}, 2020.

\bibitem{gupta}
A. Gupta et al., ``On the utility of power spectral techniques with feature selection techniques for effective mental task classification in noninvasive BCI," \textit{in IEEE Transactions on Systems, Man, and Cybernetics: Systems}, pp. 1-13, 2019.

\bibitem{wang}
Q. Wang and O. Sourina, ``Real-time mental arithmetic task recognition from EEG signals," \textit{IEEE Transactions on Neural Systems and Rehabilitation Engineering}, vol. 21, no. 2, pp. 225-232, 2013.

\bibitem{keirn}
Z.A. Keirn and J.I. Aunon, ``A new mode of communication between man and his surroundings," \textit{IEEE Transactions on Biomedical Engineering}, vol. 37, no. 12, pp. 1209-1214, 1990.

\bibitem{tim1}
Z. Pei et. al., ``EEG-based multi-class workload identification using feature fusion and selection," \textit{in IEEE Transactions on Instrumentation and Measurement}, 2020.

\bibitem{tim2}
P. Arpaia et. al., ``A wearable EEG instrument for real-time frontal asymmetry monitoring in worker stress analysis," \textit{in IEEE Transactions on Instrumentation and Measurement}, vol. 69, no. 10, pp. 8335-8343, Oct. 2020.

\bibitem{btlr}
S.R. Butler and A. Glass, ``Asymmetries in the electroencephalogram associated with cerebral dominance,” \textit{Electroencephalogr. Clin.
Neurophysiol.}, vol. 36, pp. 481-491, 1974.

\bibitem{tim3}
A. E. Alchalabi et. al., ``FOCUS: Detecting ADHD patients by an EEG-based serious game," \textit{in IEEE Transactions on Instrumentation and Measurement}, vol. 67, no. 7, pp. 1512-1520, July 2018.

\bibitem{tim4}
B. Wallace et. al., ``EEG/ERP: Within episodic assessment framework for cognition," \textit{in IEEE Transactions on Instrumentation and Measurement}, vol. 66, no. 10, pp. 2525-2534, Oct. 2017.

\bibitem{shargie}
F.M. Al-Shargie, et al., ``Mental stress quantification using EEG signals," \textit{International Conference for Innovation in Biomedical Engineering and Life Sciences}, Springer, Singapore, 2015.

\bibitem{zhang}
Y. Zhang et. al., ``Combined feature extraction method for classification of EEG signals," \textit{Neural Computing and Applications}, vol. 28, no. 11, pp. 3153-3161, 2017.

\bibitem{dutta}
S. Dutta et. al., ``Automated classification of non-motor mental task in electroencephalogram based brain-computer interface using multivariate autoregressive model in the intrinsic mode function domain," \textit{Biomedical Signal Processing and Control}, vol. 43, pp. 174-182, 2018.

\bibitem{noshadi}
S. Noshadi et. al., ``Selection of an efficient feature space for EEG-based mental task discrimination," \textit{Biocybernetics and Biomedical Engineering}, vol. 34, no. 3, pp. 159-168, 2014.

\bibitem{bash}
P. Bashivan, G. M. Bidelman, and M. Yeasin, ``Spectrotemporal dynamics of the EEG during working memory encoding and maintenance predicts
individual behavioral capacity,” \emph{European Journal of Neuroscience}, vol. 40, no. 12, pp. 3774–3784, 2014.

\bibitem{SatijaSensor2019}
M. Saini, Payal, U. Satija,``An effective and robust framework for ocular artifact removal from single-channel EEG signal based on variational mode decomposition", \emph{IEEE Sensor J.}, vo. 20, no. 1, pp.369-376, 2019.

\bibitem{SatijaIET2020}
M. Saini, U. Satija, M.D. Upadhayay, ``Effective automated method for detection and suppression of muscle artefacts from single-channel EEG signal," \emph{IET Healthcare Technology Letters}, vol. 7, no. 2, pp. 35-40, 2020.

\bibitem{confr}
A. Qayyum et. al., ``Classification of EEG learning and resting states using 1D-convolutional neural network for cognitive load assesment," \textit{2018 IEEE Student Conference on Research and Development (SCOReD)}, Selangor, Malaysia, pp. 1-5, 2018.

\bibitem{ravi}
D. Ravì et al., ``Deep learning for health informatics," \textit{in IEEE Journal of Biomedical and Health Informatics}, vol. 21, no. 1, pp. 4-21, Jan. 2017.

\bibitem{jiao}
Z. Jiao et. al., ``Deep convolutional neural networks for mental load classification based on EEG data," \textit{Pattern Recognition}, vol. 76, pp. 582-595, 2018.

\bibitem{arxv}
S. Kiranyaz et al., ``1D convolutional neural networks and applications: a survey," \textit{arXiv preprint arXiv:1905.03554}, 2019.

\bibitem{ander}
Anderson et. al., ``Classification of EEG signals from four subjects during five mental tasks," \textit{Solving Engineering Problems with Neural Networks: Proceedings of the Conference on Engineering Applications in Neural Networks (EANN-96)}, Turkey, 1996.

\bibitem{lin}
C.J. Lin and M.H. Hsieh, ``Classification of mental task from EEG data using neural networks based on particle swarm optimization," \textit{Neurocomputing}, vol. 72, no. 4-6, pp. 1121-1130, 2009.

\bibitem{acmdl}
Z.A.A. Alyasseri et. al., ``The effects of EEG feature extraction using multi-wavelet decomposition for mental tasks classification," \textit{In Proceedings of the International Conference on Information and Communication Technology}, pp. 139-146, April, 2019.

\bibitem{netbio}
R.S. Anand, Gaurav, and V. Kumar, ``EEG-metric based mental stress detection," \textit{Network Biology}, vol. 8, no. 1, pp. 25-34, 2018.

\bibitem{guo}
L. Guo et.al., ``Classification of mental task from EEG signals using immune feature weighted support vector machines," \textit{IEEE Transactions on Magnetics}, vol. 47, no.5, pp. 866-869, 2010.

\bibitem{chen}
P. Zhang et. al., ``Learning spatial–spectral–temporal EEG features with recurrent 3D convolutional neural networks for cross-task mental workload assessment," \textit{in IEEE Transactions on Neural Systems and Rehabilitation Engineering}, vol. 27, no. 1, pp. 31-42, Jan. 2019.

\bibitem{craik}
A. Craik et. al., ``Deep learning for electroencephalogram (EEG) classification tasks: a review," \textit{Journal of Neural Engineering}, vol. 16, no. 3, 2019.

\bibitem{yuan}
Y. Yuan et. al., ``A multi-view deep learning framework for EEG seizure detection," \textit{in IEEE Journal of Biomedical and Health Informatics}, vol. 23, no. 1, pp. 83-94, Jan. 2019.

\bibitem{spl2}
D. Zhang et. al., ``A convolutional recurrent attention model for subject-independent EEG signal analysis," \textit{in IEEE Signal Processing Letters}, vol. 26, no. 5, pp. 715-719, May 2019.

\bibitem{emot}
J.X. Chen et. al., ``Accurate EEG-based emotion recognition on combined features using deep convolutional neural networks," \textit{in IEEE Access}, vol. 7, pp. 44317-44328, 2019.

\bibitem{arxv1}
Z. Bai, Y. Ruizhi, and L. Youzhi, ``Mental task classification using electroencephalogram signal," \textit{arXiv preprint arXiv:1910.03023}, 2019.

\bibitem{eegmat}
I. Zyma et. al., ``Electroencephalograms during mental arithmetic task performance," \textit{Data 4}, vol. 4, no. 14, 2019.

\bibitem{physion}
A, Goldberger et. al., ``PhysioBank, physioToolkit, and physioNet: components of a new research resource for complex physiologic signals", \textit{Circulation}, vol. 101, no. 23, pp. 215-220, 2003.

\bibitem{neuro}
E. W. Nugroho and B. Harnadi, ``The method of integrating virtual reality with brainwave sensor for an interactive math's game," \textit{2019 16th International Joint Conference on Computer Science and Software Engineering (JCSSE)}, Chonburi, Thailand, pp. 359-363, 2019.

\bibitem{SatijaIoT}
U. Satija, B. Ramkumar, M. S. Manikandan, ``Real-Time signal quality-aware ECG telemetry system for IoT-based health care monitoring," \emph{IEEE Internet of Things Journal}, vol. 4, no. 3, pp. 815-823, June 2017.

\bibitem{SatijaJBHI}
U. Satija, B. Ramkumar, M. S. Manikandan, ``Automated ECG noise detection and classification system for unsupervised healthcare monitoring," \emph{IEEE Journal of Biomedical and Health Informatics}, vol. 22, no. 3, pp. 722 - 732, May 2017. I.F.-3.45.

\bibitem{croce}
P. Croce et.al., ``Deep convolutional neural networks for feature-less automatic classification of independent components in multi-channel electrophysiological brain recordings," \textit{in IEEE Transactions on Biomedical Engineering}, vol. 66, no. 8, pp. 2372-2380, Aug. 2019.

\bibitem{access}
D. Peng et. al., ``A novel deeper one-dimensional CNN with residual learning for fault diagnosis of wheel set bearings in high-speed trains," \textit{in IEEE Access}, vol. 7, pp. 10278-10293, 2019.

\bibitem{prez}
M. Pérez-Enciso et. al., ``A guide on deep learning for complex trait genomic prediction," \textit{Genes}, vol. 10, no. 7, p. 553, 2019.

\bibitem{wu}
H. Wu, and X Gu, ``Towards dropout training for convolutional neural networks," \textit{Neural Networks}, vol. 71, pp. 1-10, 2015.

\bibitem{elfw}
S. Elfwing et. al., ``Sigmoid-weighted linear units for neural network function approximation in reinforcement learning," \textit{Neural Networks}, vol. 107, pp. 3-11, 2018.

\bibitem{hanin}
B. Hanin and D. Rolnick, ``How to start training: The effect of initialization and architecture," \textit{In Advances in Neural Information Processing Systems}, vol. 31, pp. 571-581, 2018.

\bibitem{ig}
I. Goodfellow et. al., ``Deep learning," \textit{MIT press}, 2016.

\bibitem{mendeley}
M.A. Klados and P.D. Bamidis, ``A semi-simulated EEG/EOG dataset for the comparison of EOG artifact rejection techniques", \emph{Data in Brief}, vol. 8, pp. 1004-1006, 2016.

\bibitem{ccam}
W.D. Clercq et. al., ``Canonical correlation analysis applied to remove muscle artifacts from the electroencephalogram," \emph{in IEEE Transactions on Biomedical Engineering}, vol. 53, no. 12, pp. 2583-2587, Nov. 2006.
%

%

\end{thebibliography}
\end{document}